\documentclass[11pt]{article}
\usepackage{xspace}
\usepackage{graphicx}
\usepackage{amsmath}
\usepackage{amssymb}
\usepackage{color}

\definecolor{Red}{rgb}{1,0,0}
\definecolor{Green}{rgb}{0,1,0}
\definecolor{Blue}{rgb}{0,0,1}
\definecolor{Black}{rgb}{0,0,0}

\def\beq{\begin{equation}}
\def\eeq#1{\label{#1}\end{equation}}
\def\eeqn{\end{equation}}

\def\beqa{\begin{eqnarray}}
\def\eeqa#1{\label{#1}\end{eqnarray}}
\def\eeqan{\end{eqnarray}}

\let\bar=\overbar

\def\Dslash{\not{\hbox{\kern-4pt $D$}}}
\def\dslash{\not{\hbox{\kern-2pt $\del$}}}

\def\msb{{\bar{\ssstyle M \kern -1pt S}}}

\def\Title#1{\begin{center} {\Large {\bf #1} } \end{center}}

\begin{document}

\Title{Unbinned halo-independent methods for emerging dark matter signals}

\bigskip\bigskip


\begin{raggedright}  

Yonatan Kahn\index{Kahn, Y.}, {\it Massachusetts Institute of Technology}\\

\bigskip
\end{raggedright}


{\small
\begin{flushleft}
\emph{To appear in the proceedings of the Interplay between Particle and Astroparticle Physics workshop, 18 -- 22 August, 2014, held at Queen Mary University of London, UK.}
\end{flushleft}
}

\section{Introduction}

By now the existence of dark matter (DM) has been unambiguously established by multiple astrophysical measurements. Galactic rotation curves which flatten out at large radius imply the existence of galactic DM halos (see e.g.\ \cite{vanAlbada:1984js}). N-body simulations such as Via Lactea \cite{Kuhlen:2008qj} show that DM with only gravitational interactions forms structures and substructures. Cosmic microwave background measurements from the Planck satellite \cite{Ade:2013zuv} allow one to extract cosmological parameters to extremely high precision, including the component of DM in the energy budget of the universe, 26.8\%. There are even constraints on the DM self-interaction cross section, $\sigma/m < 1.3 \ \textrm{barn}/\textrm{GeV}$, from observations of the Bullet Cluster \cite{Markevitch:2003at}.

However, despite these advances, many properties of dark matter remain unknown. Its mass could lie anywhere from sub-eV to $10^{13} \ \textrm{GeV}$, with many well-motivated candidates located at all mass scales. Its non-gravitational interactions with visible matter could be elastic or inelastic, proceed through a light or heavy mediator, or it could have no such interactions. In analogy to the complexity of the Standard Model in the visible sector, there could be an entire dark sector with multiple states, gauge groups, self-interactions, and decays. Finally, we have no direct measurements of the local DM velocity distribution in our own galactic halo. A common assumption is a Maxwellian distribution, but N-body simulations suggest deviations from this distribution \cite{Kuhlen:2009vh}, which can have a strong effect on direct detection experiments.

Given our uncertainties about the properties of dark matter, it is advantageous to develop experiments and analysis techniques which make as few assumptions as possible about these properties. Specifying to direct detection experiments, which search for DM scattering off nuclei, the differential event rate for spin-independent scattering as a function of nuclear recoil energy is 

\begin{equation}
\label{eq:DiffRate}
\frac{dR}{d E_R} = \frac{N_A \rho_\chi \sigma_n m_n}{2 m_\chi \mu_{n\chi}^2} C_T^2 (A,Z) \int d E'_R G (E_R,E'_R) \epsilon(E'_R) F^2 (E'_R) \int^\infty_{v_{min}(E'_R)} \frac{f(\bf{v}+\bf{v}_E)}{v} d^3 v.
\end{equation} 
This expression contains input from the DM model (density $\rho_\chi$, nuclear cross section $\sigma_n$, masses $m_\chi$ and $\mu_{n \chi}$), the detector properties (target-dependent coherent scattering enhancement $C_T^2(A,Z)$, detector resolution function $G(E_R, E'_R)$, and detector efficiency $\epsilon(E'_R$)), nuclear physics (nuclear form factor $F^2(E'_R)$), and the halo model (DM velocity distribution $f(\bf{v} + \bf{v}_E)$, where $\bf{v}_E$ is the velocity of the Earth). In addition, the lower limit $v_{min}(E'_R)$ of the halo integral, which is the minimum dark matter velocity required to provoke a nuclear recoil $E'_R$, depends on the kinematics of the DM model. The traditional method for analyzing direct detection experiments is to choose a dark matter model and a halo model (for example, a Maxwellian velocity distribution), and present exclusion limits or preferred regions in $m_\chi - \sigma_n$ space. However, an alternate, ``halo-independent'' analysis \cite{Fox:2010bz,Fox:2010bu} is possible: rather than choosing a halo model, one can simply change variables and present exclusion limits or preferred regions in $v_{min} - g(v_{min})$ space, where
\begin{equation}
\label{eq:gvmin}
g(v_{min}) = \int^\infty_{v_{min}} \frac{f(\bf{v}+\bf{v}_E)}{v} d^3 v
\end{equation}
is the halo integral written as a function of its lower limit. This requires no assumptions about the DM halo, and makes it easy to compare multiple experiments, because two experiments with different nuclear targets may have overlapping ranges of $v_{min}$ even if they have non-overlapping ranges of $E_R$. 

The present null results from direct detection experiments such as XENON100 \cite{Aprile:2012nq} and LUX \cite{Akerib:2013tjd} suggest that an emerging dark matter signal will likely consist of only a handful of events, and thus it is advantageous to keep as much information about each event as possible. Direct detection experiments have extremely low backgrounds and excellent energy resolution, so one should avoid binning the data and use unbinned analysis techniques. Furthermore, one should test a potential signal against as many DM scattering kinematics as possible, in particular not just elastic scattering. We will focus here on extending the methods of \cite{Fox:2010bz} to unbinned data, based on \cite{Fox:2014kua} with Patrick Fox and Matthew McCullough, as well as a further generalization to inelastic kinematics \cite{KahnToAppear}.

\section{Exploiting monotonicity}

Powerful consistency conditions on a putative dark matter signal can be derived from the simple observation that the integrand of Eq.\ (\ref{eq:gvmin}) is positive-definite \cite{Fox:2010bz,Fox:2010bu}, and thus by the Fundamental Theorem of Calculus, $g(v_{min})$ is a monotonically decreasing function of $v_{min}$. As mentioned in the introduction, one must specify a model of DM scattering in order to fix $v_{min}(E_R)$ and relate the halo integral to an experiment which measures nuclear recoil. For now we will focus on the case of elastic scattering:
\begin{equation}
\label{eq:vminelastic}
v_{min}(E_R) = \sqrt{\frac{m_N E_R}{2\mu^2_{N\chi}}},
\end{equation}
where $m_N$ is the mass of the nuclear target and $\mu_{N\chi}$ is the DM-nucleus reduced mass.

\begin{figure}[!ht]
\begin{center}
\includegraphics[width=0.5\columnwidth]{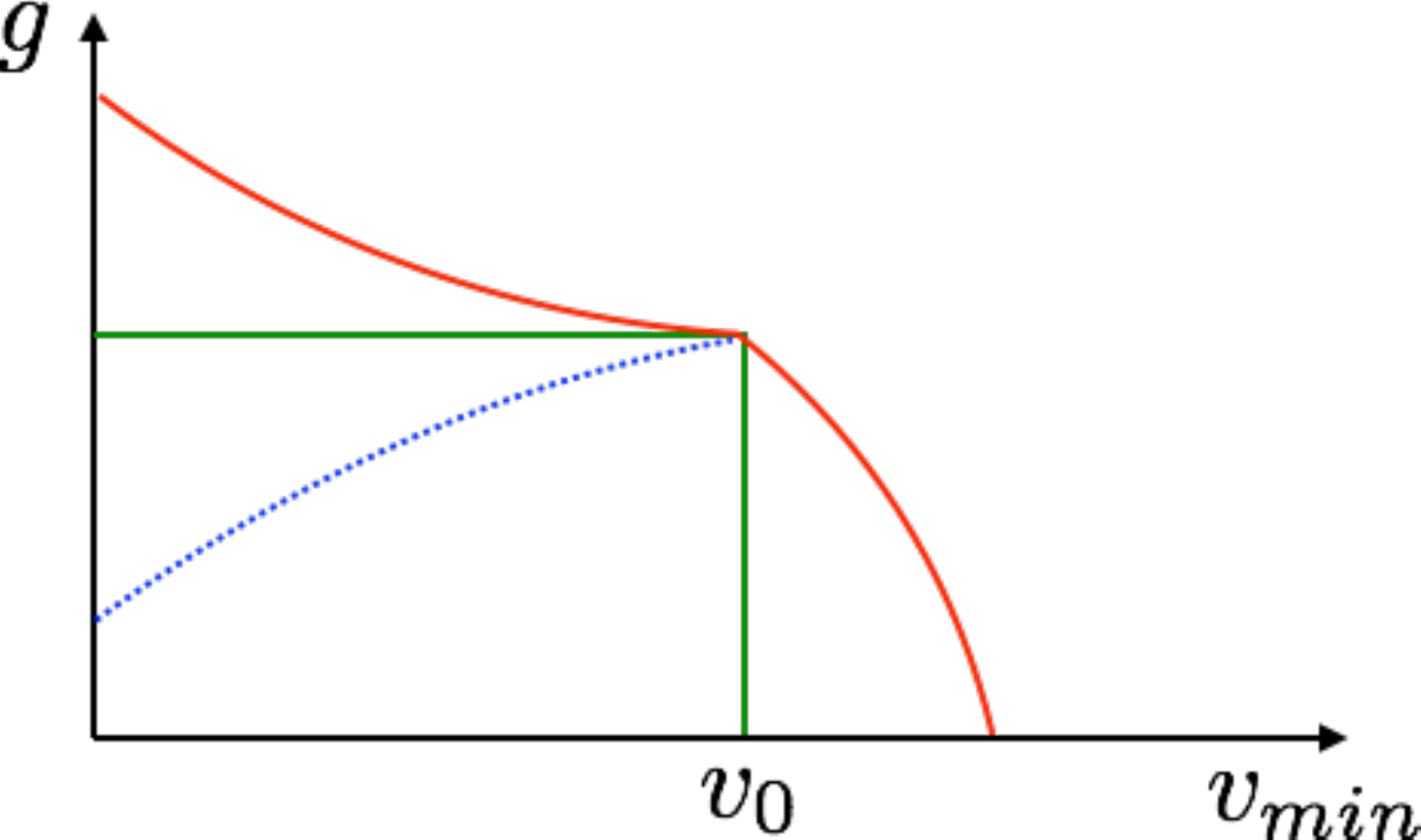}
\caption{Cartoon of some choices of $g(v_{min})$: for a given $v_0$, green is most conservative, red is allowed but gives more total events, and blue is inconsistent because it violates the monotonicity constraint.}
\label{fig:gchoices}
\end{center}
\end{figure}

If an experiment obtains a null results, one can set exclusion limits in $v_{min}-g(v_{min})$ space. Eq.\ (\ref{eq:vminelastic}) gives a 1-to-1 mapping between a DM velocity $v_0$ and a nuclear recoil energy $E_0$. The monotonicity constraint implies that at each $v_0$, the most conservative choice of halo integral is a step function,
\begin{equation}
\label{eq:gstep}
g(v_{min}) = g(v_0) \theta(v_0 - v_{min}),
\end{equation}
since this choice for the form of $g(v_{min})$ minimizes the value of the halo integral, and hence the total number of events expected from Eq.\ (\ref{eq:DiffRate}). This will then lead to the weakest possible exclusion limit for the halo integral $g(v_0)$ evaluated at $v_{min} = v_0$. Figure \ref{fig:gchoices} shows three choices for $g(v_{min})$: the green curve is the step function (\ref{eq:gstep}), the red curve is another halo which satisfies the monotonicity constraint but gives more total events, while the blue curve does not satisfy the monotonicity constraint and hence is inconsistent with a DM halo. To build up an exclusion curve, one simply plugs the step function form (\ref{eq:gstep}) into the rate equation (\ref{eq:DiffRate}), sets limits on the height of the step $g(v_0)$ based on the null result, and repeats for all $v_0$ in the range of the experiment, using (\ref{eq:vminelastic}) to map a range of $E_R$ into a range of $v_{min}$. Similarly, if an experiment sees a positive signal, one can map the event rate in an energy range $[E_1, E_2]$ to a rate in the $v_{min}$ range $[v_1, v_2]$, and use maximum likelihood techniques to determine the preferred values for $g(v_{min})$ in each range.

\begin{figure}[!ht]
\begin{center}
\includegraphics[width=0.5\columnwidth]{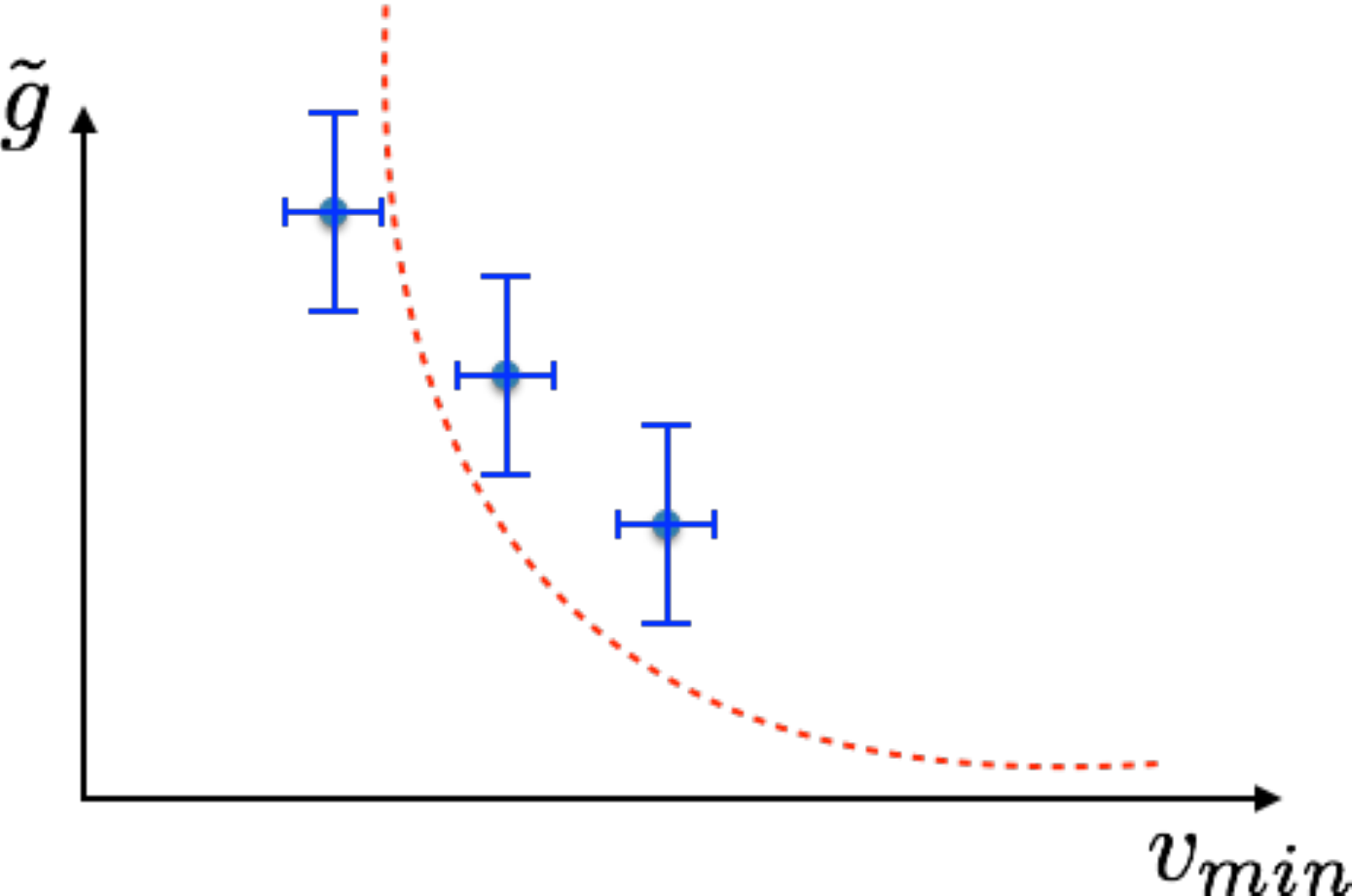}
\caption{Cartoon of an exclusion limit from one experiment (red dashed) and a signal from another experiment (blue points). The two signal points lying above the red exclusion curve are excluded for all possible DM halos, at the given confidence level.}
\label{fig:compare}
\end{center}
\end{figure}

In fact, one can easily compare a null result from one experiment with a potential signal from another experiment using these techniques \cite{Frandsen:2011gi}. By rescaling the halo integral,
\begin{equation}
\label{eq:gtilde}
\tilde{g} (v_{min}) = \frac{\rho_\chi \sigma_n}{m_\chi} g(v_{min}),
\end{equation}
the factors common to all experiments (but dependent on the DM model) are included in $\tilde{g}$, and the preferred values and exclusion limits can be placed on the same plot. In particular, no assumptions need be made about the local DM density $\rho_\chi$ or the scattering cross section $\sigma_n$, since both simply rescale $\tilde{g}$. A cartoon example is shown in Figure \ref{fig:compare}, where the red dashed curve represents an exclusion limit, and the three blue points represent positive signals, with bin widths given by the horizontal error bars and confidence intervals given by the vertical error bars. On such a plot, a signal point lying above the red exclusion limit is excluded for \emph{all} halos, since for each $v_{min}$ the most conservative possible halo was used to build up the exclusion curve. Furthermore, if the signal points did not satisfy the monotonicity condition on $\tilde{g}$, they would be inconsistent with a DM interpretation independent of any exclusion limits. This presentation in $v_{min}-\tilde{g}(v_{min})$ space is nicely complementary to the usual $m_\chi - \sigma_n$ plots. For other work on halo-independent methods, see \cite{Gondolo:2012rs,HerreroGarcia:2012fu,Bozorgnia:2013hsa,Peter:2013aha,Feldstein:2014gza,DelNobile:2014sja}.

\section{Unbinned halo-independent methods for elastic scattering}
\label{sec:Unbinned}

\begin{figure}[!ht]
\begin{center}
\includegraphics[width=0.5\columnwidth]{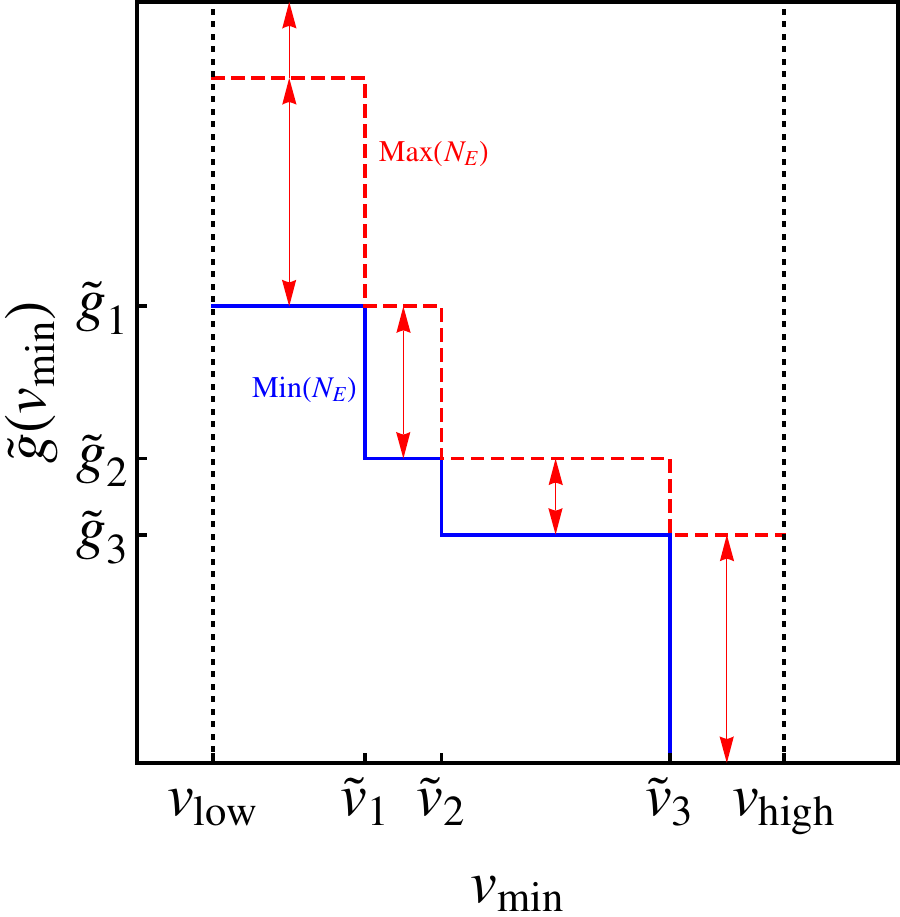}
\caption{Cartoon of best-fit (blue) and worst-fit (red) shapes for $\tilde{g} (v_{min})$. For perfect energy resolution, $\tilde{v}_i = v_{min}(E_i)$, where $E_i$ are the measured energies.}
\label{fig:AllHalo}
\end{center}
\end{figure}

To perform a halo-independent analysis on unbinned data \cite{Fox:2014kua}, one can use the extended maximum likelihood method \cite{Barlow:1990vc}:
\begin{equation}
\mathcal{L} = \frac{e^{-N_{E}}}{N_{O}!} \prod^{N_{O}}_{i=1} \frac{dR_T}{d E_R} \bigg|_{E_R=E_i},
\end{equation}
where $dR_T/dE_R$ is the total differential event rate (signal plus background), and 
\begin{equation}
N_{E} = \int_{E_{min}}^{E_{max}}  \frac{dR_T}{d E_R} d E_R.
\end{equation}
is the total number of events expected for a given set of parameters, for $E_R \in [E_{min}, E_{max}]$. Note that $\mathcal{L}$ penalizes against more expected events. Indeed,
\begin{equation}
N_{E} \propto \int dE'_R\, \tilde{g}(v_{min}(E'_R)) \propto \int dv_{min}\, \tilde{g}(v_{min}),
\end{equation}
so for fixed differential event rates, $\mathcal{L}$ is maximized when the area under $\tilde{g}(v_{min})$ is minimized. Taking into account the monotonicity constraint, this implies that the best-fit form of $\tilde{g}(v_{min})$ is a sum of step functions, as shown in Figure \ref{fig:AllHalo} (blue curve). Here, the positions of the steps $\tilde{v}_i$ correspond to $v_{min}(E_i)$ in the case of perfect energy resolution, $G(E_R, E_R') = \delta(E_R - E_R')$, but can shift for finite energy resolution, as will be discussed below. The worst-fit form passing through the same points $(\tilde{v}_i, \tilde{g}_i)$ (red curve) is also a sum of step functions, but one where the height of the first step is taken to infinity, which maximizes rather than minimizes $N_E$.

For finite energy resolution, 
\begin{equation}
\frac{dR_T}{dE_R}\bigg|_{E_i} = \frac{dR_{BG}}{d E_R}\bigg|_{E_i} + \frac{N_A m_n}{2  \mu_{n\chi}^2} C_T^2 (A,Z) \int d E'_R G (E_i,E'_R) \epsilon(E'_R) F^2 (E'_R) \tilde{g}(v_{min} (E'_R)),
\end{equation}
and the expression for the total differential event rate at $E_i$ depends on an integral over the entire function $\tilde{g}(v_{min} (E'_R))$. One may worry that the simple step-function form discussed above no longer maximizes the likelihood, but we prove in \cite{Fox:2014kua} using variational techniques that this is not the case; the only modification to the arguments above is that the positions of the steps $\tilde{v}_i$ should be allowed to float. Thus, maximizing the extended likelihood function $\mathcal{L}$ for $N_O$ events requires only a $2N_{O}$-parameter numerical maximization over the heights $\tilde{g}_i$ and positions $\tilde{v}_i$ of the steps.

\begin{figure}[!ht]
\begin{center}
\includegraphics[width=0.5\columnwidth]{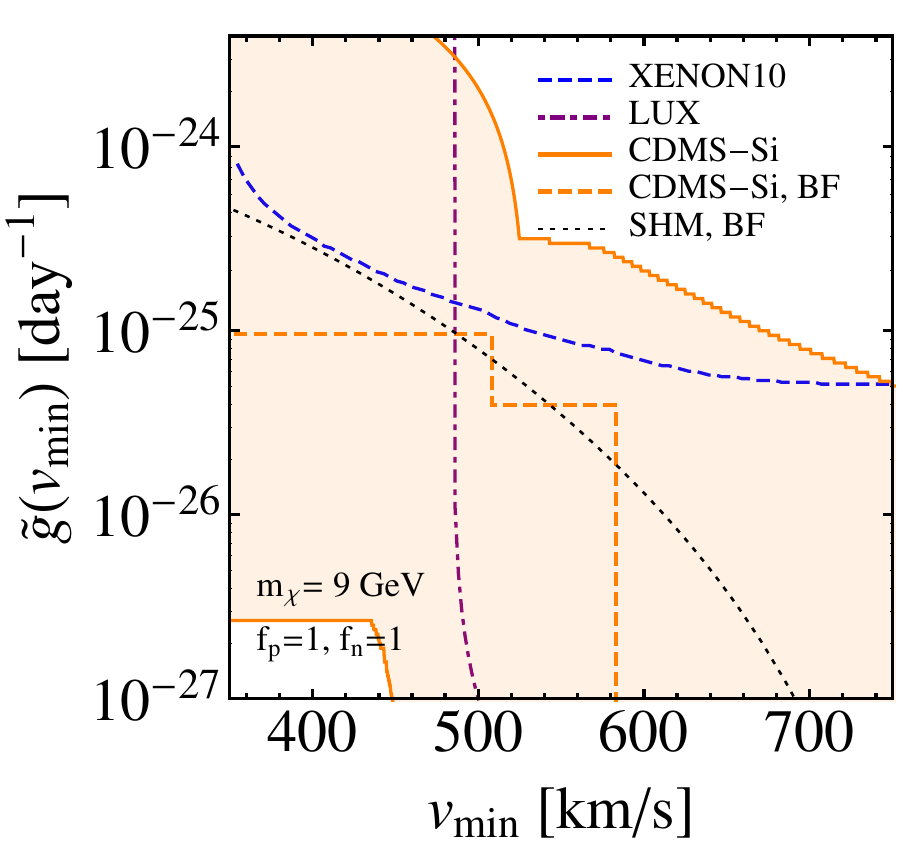}
\caption{Halo-independent interpretation of the CDMS-Si events versus constraints from XENON10 and LUX for $m_\chi = 9 \ \textrm{GeV}$ assuming elastic, spin-independent scattering with equal couplings to protons and neutrons. The preferred envelope and constraints are both calculated at $90\%$ confidence.  The best-fit halo is inconsistent with the LUX results and only a small section of the lower boundary of the preferred halo envelope for CDMS-Si is compatible with the null LUX results. The curve for the SHM is also shown, giving a good fit to the CDMS-Si data as well as a curve for the best-fit halo which minimizes the extended likelihood.}
\label{fig:CDMSvsLUX}
\end{center}
\end{figure}

As an example of this method, we compare the three events from CDMS-Si \cite{Agnese:2013rvf} to exclusion limits from LUX \cite{Akerib:2013tjd} and XENON10 \cite{Angle:2011th}. The results are shown in Figure \ref{fig:CDMSvsLUX} for $m_\chi = 9 \ \textrm{GeV}$ and $f_p = f_n$, i.e.\ identical couplings to protons and neutrons. The limits from LUX exclude both the standard halo model (black dashed) and the best-fit halo for the three CDMS events (orange dashed). Note that despite the three events, there are only two steps, because the best-fit form for $\tilde{g}(v_{min})$ prefers two adjacent steps to have the same height. Due to the finite-resolution effects mentioned above, at a given confidence level the preferred region for CDMS is an envelope (shaded orange), rather than a curve with vertical error bars. If a portion of the bottom of the envelope were excluded, that would imply that the CDMS events were excluded for all halos. As it stands, neither exclusion curve crosses the bottom of the envelope, meaning that though there is some tension between LUX and CDMS, a DM interpretation of the CDMS events cannot be ruled out in a halo-independent fashion.

To make a plot like Figure \ref{fig:CDMSvsLUX}, one must choose a fiducial value of $m_\chi$, in order to calculate the rate and perform the halo-independent analysis. However, there is a simple rescaling which can be applied to such a plot to compare experiments for any value of $m_\chi$. The rate equation (\ref{eq:DiffRate}) and the expression for $v_{min}$ in (\ref{eq:vminelastic}) imply that under a change $m_\chi \to m'_\chi$, the axes of Figure \ref{fig:CDMSvsLUX} transform as follows:
\begin{align}
v'_{min}(E_R)&  = \frac{\mu_{N\chi}}{\mu_{N\chi'}}v_{min}(E_R), \\
\tilde{g}' & = \frac{\mu^2_{n\chi'}}{\mu^2_{n\chi}} \tilde{g}.
\end{align}
Thus to evaluate exclusions and best-fit regions for a different DM mass, one need not perform the entire analysis again from scratch, but rather one needs only to rescale each experiment by the constant factors given above. In this sense, one halo-independent plot contains all the information necessary to compare multiple experiments, for any DM mass.

\section{Generalization for a more complex dark sector}

The method of Sec.\ \ref{sec:Unbinned} can be extended in a straightforward way to more general DM-nucleus scattering kinematics \cite{KahnToAppear}. The principal difference is that $v_{min}$ is no longer required to be a 1-to-1 function of $E_R$. In particular, for inelastic or exothermic scattering,
\begin{equation}
v_{min}(E_R) = \sqrt{\frac{1}{2 m_N E_R}} \left | \frac{m_N E_R}{\mu_{N\chi}} + \delta \right |,
\end{equation}
where $\delta/2$ is the mass splitting between the incoming and outgoing DM states. Positive $\delta$ corresponds to inelastic scattering, and negative $\delta$ to exothermic scattering. These cases have been treated in \cite{Bozorgnia:2013hsa} in the case of binned data; here we focused on unbinned approaches.

\begin{figure}[!ht]
\begin{center}
\includegraphics[width=0.5\columnwidth]{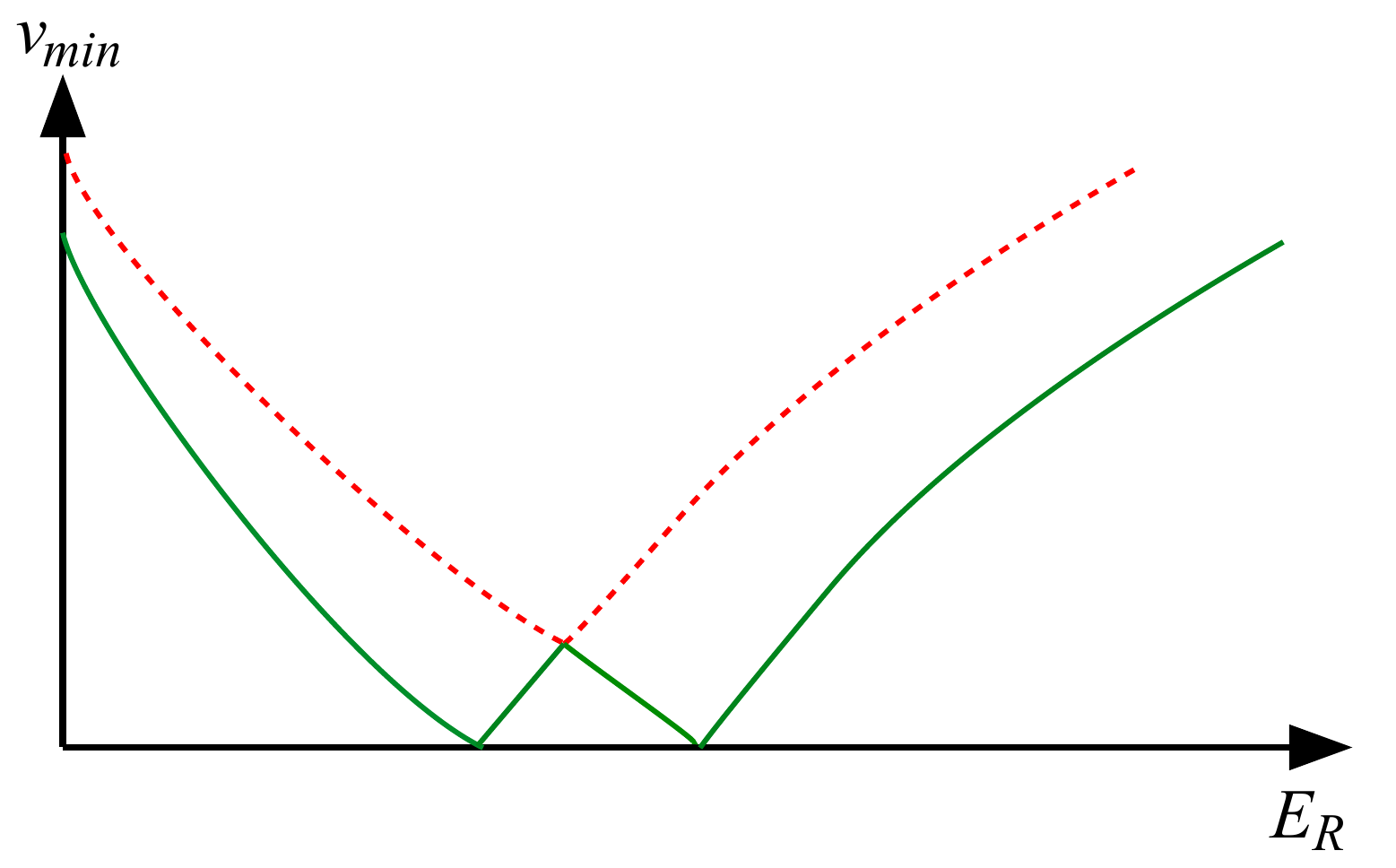}
\caption{Cartoon of the $v_{min}(E_R)$ curves for a three-level system with two possible exothermic down-scatterings. $v_{min}^{+}$ is shown in red (dashed lines), and $v_{min}^{-}$ in green (solid lines). Note that while the $v_{min}(E_R)$ for each channel is a 2-to-1 function, $v_{min}^{-}$ is a 4-to-1 function for small $v_{min}$.}
\label{fig:multivmin}
\end{center}
\end{figure}

One can imagine other ways in which the effective $v_{min}$ is a many-to-one function of $E_R$. Suppose dark matter-nucleus scattering has several possible scattering channels; for example, three-level exothermic dark matter with two possible down-scatterings. Let the total inclusive scattering cross section be fixed, and the relative cross sections for the two scattering channels be $\alpha_1$ and $\alpha_2$, with $\alpha_1 + \alpha_2 = 1$. Let $v_{min, i}(E_R)$ be the $v_{min}$ associated to scattering channel $i$. Notice that because the velocity integrand $f({\bf v}+ {\bf v}_E)/v$ is positive-definite, for a given $E_R$ the smallest possible value of the velocity integral over all the scattering channels will come from the largest of the $v_{min, i}(E_R)$, and similarly the largest value of the integral will come from the smallest of the $v_{min, i}$. Thus we can bound the differential event rates from above and below in a model-independent way, as follows:
\begin{equation}
\label{eq:MultiChannelBound}
\left . \frac{dR}{dE_R} \right |_{v_{min} = v_{min}^{+}} \leq \frac{dR}{dE_R}(v_{min, i}) \leq \left . \frac{dR}{dE_R} \right |_{v_{min} = v_{min}^{-}},
\end{equation}
where
\begin{equation}
v_{min}^{+} = \max_i \{v_{min, i}(E_R)\}, \hspace{3mm} v_{min}^{-} = \min_i \{v_{min, i}(E_R)\}.
\end{equation}
The shapes of these ``composite'' functions $v_{min}^{\pm}(E_R)$ may be considerably different from the those of the individual channels, as shown in Figure \ref{fig:multivmin}. However, this example illustrates that a model-independent analysis of a system with multiple scattering channels requires techniques for many-to-1 $v_{min}$.

\begin{figure}[!ht]
\begin{center}
\includegraphics[width=0.5\columnwidth]{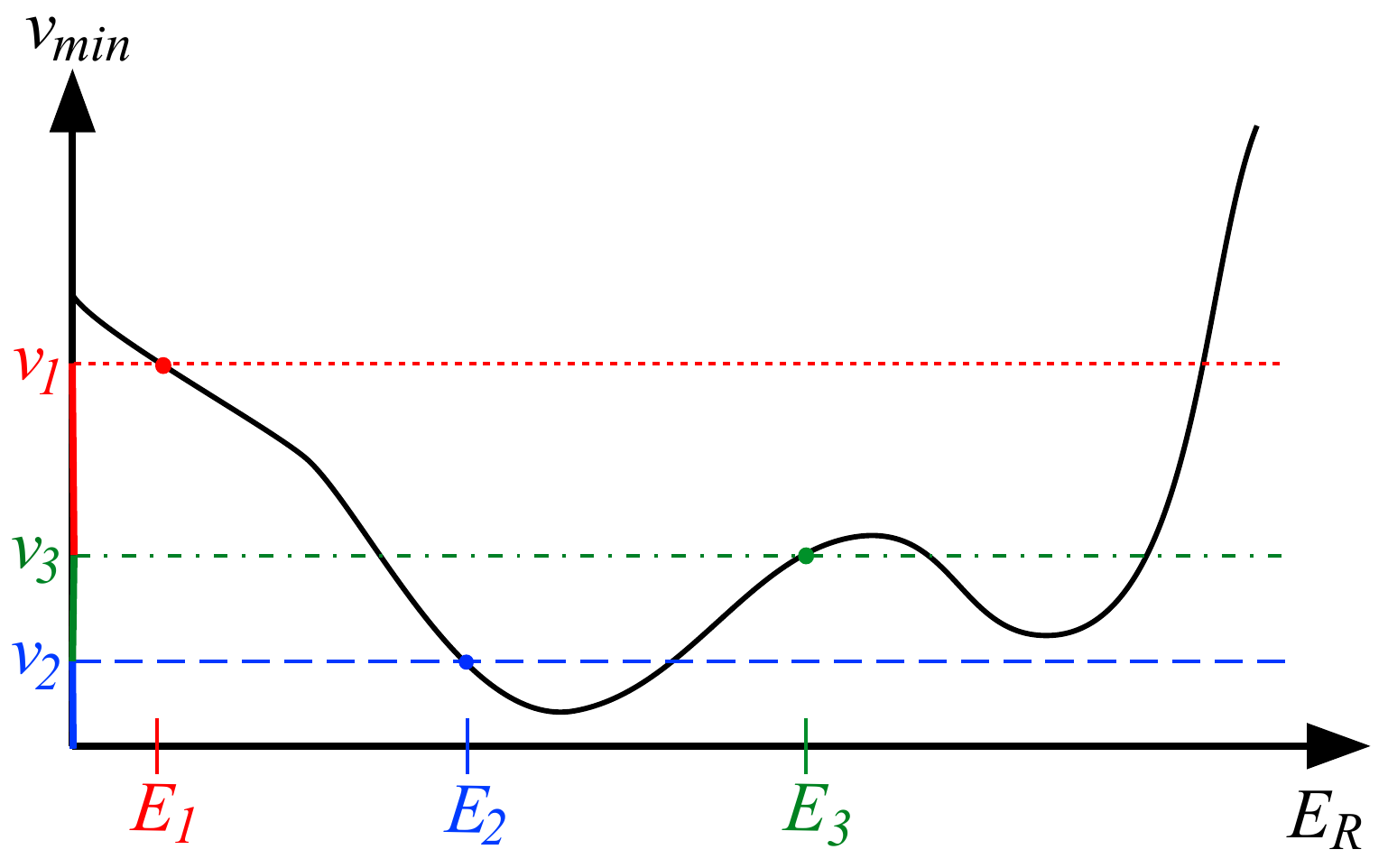}
\caption{A generic form of $v_{min}(E_R)$. $N_O = 3$ events are seen at energies $E_1$ (red), $E_2$ (blue), and $E_3$ (green). The corresponding $v_{min}$ values $v_1$, $v_2$, and $v_3$ are shown in red dotted, blue dashed, and green dot-dashed lines, respectively.}
\label{fig:vminSimple}
\end{center}
\end{figure}

An example will serve to illustrate the generalization of the unbinned halo-independent method to many-to-1 $v_{min}$. Consider the situation shown in Figure \ref{fig:vminSimple}: $N_O = 3$ events are seen at energies $E_1$, $E_2$ , and $E_3$. Since $v_{min}$ is not 1-to-1, the $v_i$ now correspond to several $E_i$ given by the intersections of the colored horizontal lines with the $v_{min}$ curve, though of course only a single $E_i$ (shown by dots) is observed.

\begin{figure}[!ht]
\begin{center}
\includegraphics[width=0.5\columnwidth]{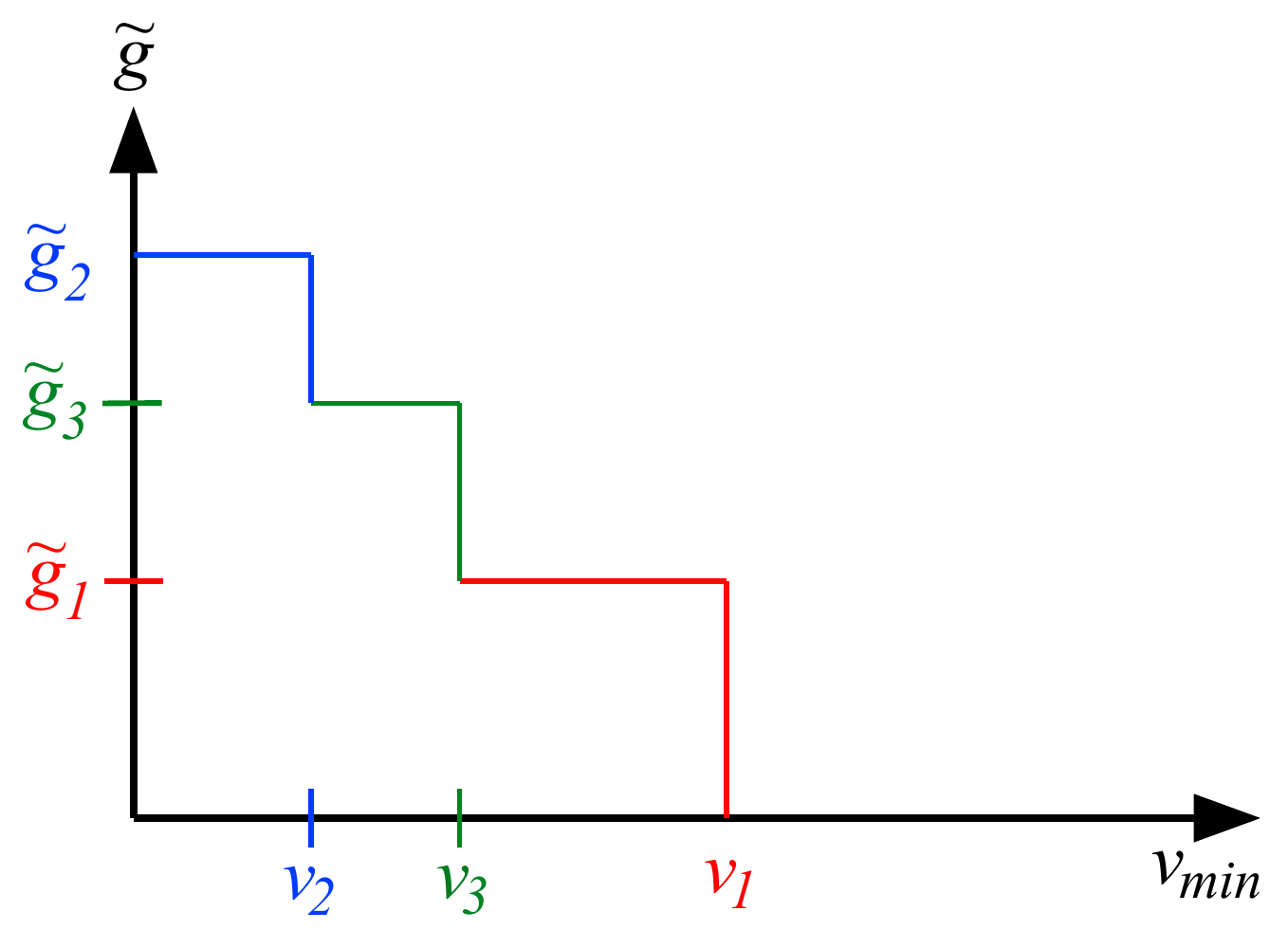}
\caption{$\tilde{g}(v_{min})$ for $E_i$ and $v_{min}(E_R)$ as given in Figure \ref{fig:vminSimple}.}
\label{fig:gtwiddleGeneral}
\end{center}
\end{figure}

\begin{figure}[!ht]
\begin{center}
\includegraphics[width=0.5\columnwidth]{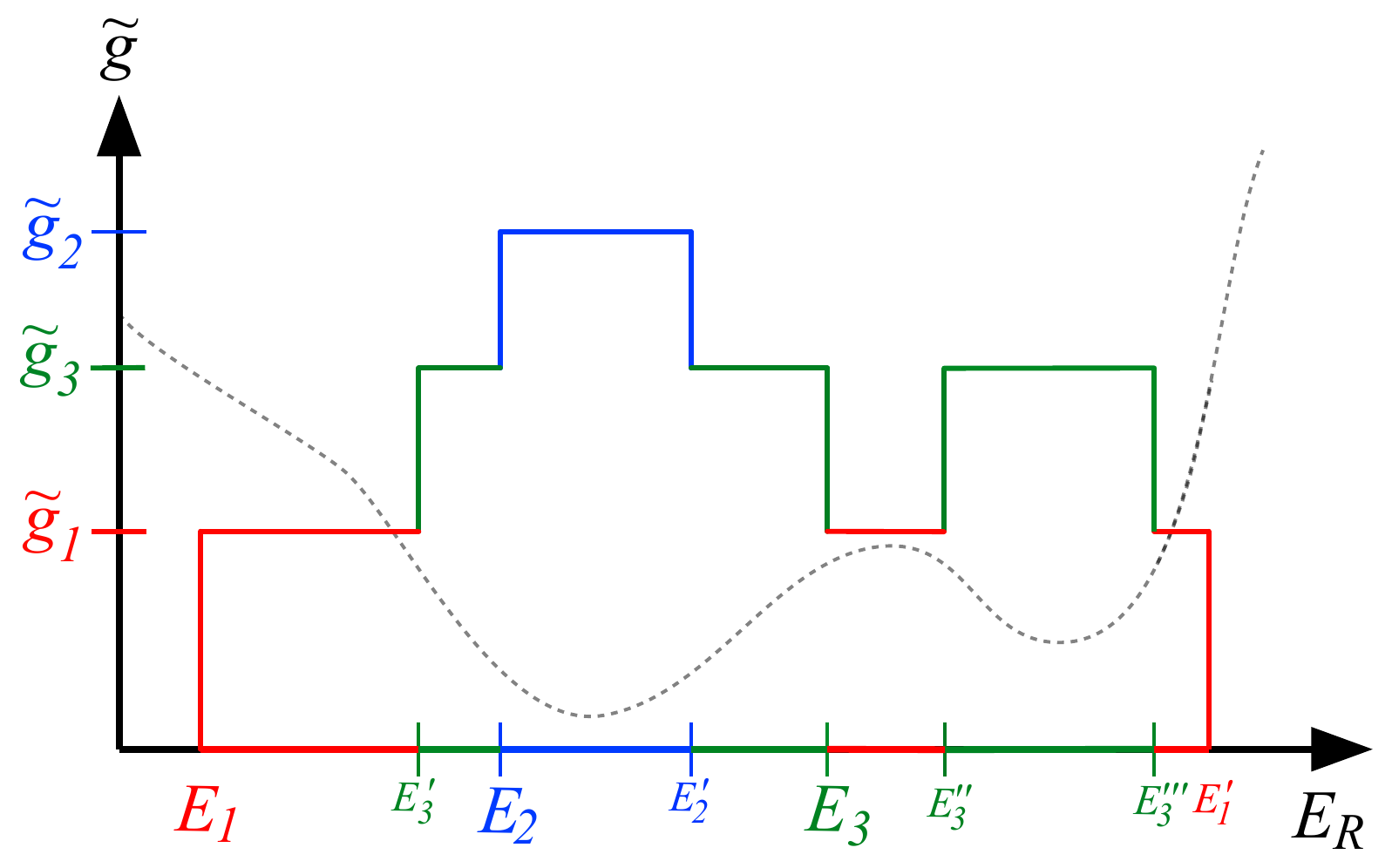}
\caption{$\tilde{g}(E_R)$ for $E_i$ and $v_{min}(E_R)$ given in Figure \ref{fig:vminSimple}. The shape of $v_{min}$ is shown as a dotted line to guide the eye.}
\label{fig:gtwiddleERGeneral}
\end{center}
\end{figure}

Analogous to the case of elastic scattering, one can prove that for sufficiently sharp energy resolution, the form of $\tilde{g}(v_{min})$ which maximizes the extended likelihood is a sum of step functions, as shown in Figure \ref{fig:gtwiddleGeneral}. Despite the many-to-1 form of $v_{min}$, for a given form of $\tilde{g}(v_{min})$ there is a unique construction of $\tilde{g}(E_R)$, shown in Figure \ref{fig:gtwiddleERGeneral}. This is essentially an ``unfolding'' of the graph of Figure \ref{fig:gtwiddleGeneral}, using the fact that $\tilde{g} = g_i$ for all $E_R$ corresponding to the same $v_{min}$. Of course, $\tilde{g}(E_R)$ is no longer monotonic, but this is hardly surprising because $v_{min}(E_R)$ is not monotonic. Indeed, $\tilde{g}$ is intrinsically a function of $v_{min}$, not $E_R$, and need not satisfy any special properties when written as a function of $E_R$. To proceed, one simply plugs this form of $\tilde{g}(E_R)$ into the rate equation (\ref{eq:DiffRate}), and computes limits or best-fit regions exactly as before.

\section{Summary}

Halo-independent methods provide a powerful tool for analyzing direct detection experiments independent of the properties of the dark matter halo. Unbinned halo-independent methods are most useful for the near term, where an emerging dark matter signal will only consist of a few events, and such methods can be extended in a straightforward way to more general dark matter scattering kinematics. While dark matter direct detection experiments are making fantastic progress, we are still ignorant about many aspects of DM, and thus it is advantageous to be as agnostic as possible about both the DM model and the halo model. We encourage experimental collaborations to present results in $v_{min} - \tilde{g}(v_{min})$ space as well as $m_\chi - \sigma_n$ space; the simple scaling with $m_\chi$ ensures that no information is lost with a fiducial choice of $m_\chi$.

\bigskip
\section{Acknowledgments}

I am grateful to my collaborators Patrick Fox and Matthew McCullough for introducing me to halo-independent techniques, as well as to Jesse Thaler for guidance regarding this and other projects. I thank Wolfgang Rau of the CDMS collaboration, Chamkaur Garg of the LUX collaboration, and especially Andrew Brown of the XENON collaboration for stimulating discussions regarding this work. I thank the organizers of IPA 2014 for an excellent conference. This work was supported in part by an NSF Graduate Fellowship.

%
%

%
%
%
%
 
\end{document}